\documentclass[
twocolumn,
10pt,
superscriptaddress,
amsmath,
amssymb,
aps,
prx,
showpacs,
floatfix,
]{revtex4-2}

\usepackage{amsfonts}
\usepackage{graphicx}
\usepackage{dcolumn}
\usepackage{bm}
\usepackage{hyperref}
\usepackage{cleveref}
\usepackage{physics}
\usepackage{xcolor}
\usepackage{mathrsfs}
\usepackage{mathtools}
\usepackage{comment}

\newcommand{\E}[0]{\mathbb{E}}
\newcommand{\V}[0]{\mathbb{V}}

\newcommand{\cov}[0]{\operatorname{Cov}}

\begin{document}
\title{Looking elsewhere: improving variational Monte Carlo gradients by importance sampling}
\date{\today}

\author{Antoine Misery}
\affiliation{CPHT, CNRS, Ecole Polytechnique, Institut Polytechnique de Paris, 91120 Palaiseau, France}
\affiliation{Collège de France, Université PSL, 11 place Marcelin Berthelot, 75005 Paris, France}
\affiliation{Inria Paris-Saclay, Bâtiment Alan Turing, 1, rue Honoré d’Estienne d’Orves – 91120 Palaiseau}
\affiliation{LIX, CNRS, École polytechnique, Institut Polytechnique de Paris, 91120 Palaiseau, France}

\author{Luca Gravina}
\affiliation{Institute of Physics, Ecole Polytechnique Fédérale de Lausanne (EPFL), CH-1015 Lausanne, Switzerland}
\affiliation{Center for Quantum Science and Engineering, Ecole Polytechnique
Fédérale de Lausanne (EPFL), CH-1015 Lausanne, Switzerland}
\author{Alessandro Santini}
\affiliation{CPHT, CNRS, Ecole Polytechnique, Institut Polytechnique de Paris, 91120 Palaiseau, France}
\affiliation{Collège de France, Université PSL, 11 place Marcelin Berthelot, 75005 Paris, France}
\affiliation{Inria Paris-Saclay, Bâtiment Alan Turing, 1, rue Honoré d’Estienne d’Orves – 91120 Palaiseau}
\affiliation{LIX, CNRS, École polytechnique, Institut Polytechnique de Paris, 91120 Palaiseau, France}

\author{Filippo Vicentini}
\affiliation{CPHT, CNRS, Ecole Polytechnique, Institut Polytechnique de Paris, 91120 Palaiseau, France}
\affiliation{Collège de France, Université PSL, 11 place Marcelin Berthelot, 75005 Paris, France}
\affiliation{Inria Paris-Saclay, Bâtiment Alan Turing, 1, rue Honoré d’Estienne d’Orves – 91120 Palaiseau}
\affiliation{LIX, CNRS, École polytechnique, Institut Polytechnique de Paris, 91120 Palaiseau, France}

\begin{abstract}
Neural-network quantum states (NQS) offer a powerful and expressive ansatz for representing quantum many-body wave functions. 
However, their training via Variational Monte Carlo (VMC) methods remains challenging. 
It is well known that some scenarios - such as sharply peaked wave functions emerging in quantum chemistry - lead to high-variance gradient estimators hindering the effectiveness of variational optimizations. In this work we investigate a systematic strategy to tackle those sampling issues by means of adaptively tuned importance sampling. 
Our approach is explicitly designed to (i) target the gradient estimator instead of the loss function, (ii) not introduce additional hyperparameters and (iii) be computationally inexpensive. 
We benchmarked our approach across the ground-state search of a wide variety of hamiltonians, including frustrated spin systems and ab-initio quantum chemistry.
We also show systematic improvements on the infidelity minimization in the context of neural projected quantum dynamics. 
Overall, our approach can reduce the computational cost of vanilla VMC considerably, up to a factor of 100x when targeting highly peaked quantum chemistry wavefunctions.
\end{abstract}

\maketitle

\section{Introduction}
\label{sec:intro}
    Variational Monte Carlo (VMC) algorithms are a powerful tool for managing the otherwise exponential complexity of quantum many-body systems.
    These methods rely on two key components:
    (i) a variational ansatz that effectively compresses the exponentially large wavefunction into a tractable set of polynomially many parameters, and (ii) Monte Carlo sampling, which reduces the computational cost of evaluating observables, loss functions, and their gradients.
    In modern applications, VMC algorithms are often combined with Neural-Network (NN) ansatze for the wavefunction, commonly referred to as Neural Quantum States (NQS) \cite{carleo_solving_2017}.
    The promise of NNs lies in their ability to efficiently represent both area-law and volume-law entangled states \cite{sharir_deep_2020} in arbitrary geometries. 
    Furthermore, networks can be engineered to encode lattice symmetries \cite{choo_symmetries_2018,Reh2023DesignSymm,Nutakki2025}, bosonic and fermionic exchange statistics \cite{luo_backflow_2019,choo_fermionic_2020,robledo_moreno_fermionic_2022}, as well as gauge symmetries in field theories \cite{Luo2021Gauge,Luo2023,Apte2024} and non-equilibrium density matrices \cite{Torlai2018_Latent,Vicentini2022_PositiveDefinite,Vicentini2019,Luo2022_AutoregOpen,Eeltink2023_open,Reh2021}.
    The \textit{expressivity} of NNs can be tuned by changing the number of parameters \cite{chen_empowering_2024,Dash2025}. 
     These parameters are typically optimized by following a stochastic estimate of the natural gradient of the objective function \cite{sorella_green_1998,Amari1998,Martens2020}.

    While the expressiveness of variational ansatze has significantly advanced since the introduction of NQS, the Monte Carlo estimators used for estimating gradients have remained largely unchanged since the early days of VMC. 
    The conventional approach draws samples from the Born probability distribution $\abs{\psi_\theta(x)}^2$ and employs these samples to empirically estimate expectation values of objective functions (including energy and fidelity), their gradients, and the metric tensor.
    Although this sampling strategy might be justified for estimating observables, its use in computing gradients of objectives and metric tensors has been challenged. Reference~\cite{sinibaldi_unbiasing_2023} demonstrated that such an approach can introduce systematic bias and lead to exploding-variance issues that can severely impair VMC performance. More recently, Ref.~\cite{gravina_neural_2024} showed that the stability and asymptotic accuracy of natural gradient descent are highly sensitive to the amount of MC noise of stochastic gradient estimates.
    

    \paragraph{Sampling issues in ground-state calculations $\qquad$}
    Sampling-related challenges are well-documented in various spin models, such as the $J_1$-$J_2$ model at high frustration \cite{chen_efficient_2023}, or the XXZ model in the anti-ferromagnetic phase \cite{park_expressive_2022}. 
    Peaked wave functions emerging from $2^\text{nd}$ quantized molecular Hamiltonians also lead to the failure of VMC and require the use of techniques such as non-stochastic sampling on a preselected subspace or sampling without replacement \cite{li_non-stochastic_2023, malyshev_neural_2024, liu_efficient_2025, li_improved_2024, cao_vision_2024}. 
    The problem is often twofold. 
    First, the structure of the wavefunction can hinder the performance of Markov Chain Monte Carlo (MCMC) sampling, leading to poor mixing times and thus a smaller effective number of samples. 
    Second, the intrinsic variance of the gradient estimated with the Born distribution can be prohibitively large. 
    These two issues can be addressed independently.
    While for the first one, autoregressive sampling strategies \cite{hibat-allah_recurrent_2020, sharir_deep_2020, malyshev_autoregressive_2023, zhao_scalable_2023} or improved MCMC algorithms \cite{bagrov_kinetic_2021, bravyi_rapidly_2023} have been developed, the second one remains largely unexplored and will be the focus of this manuscript.
    
    The brute-force approach to increase the signal to noise ratio is to increase the number of samples, but this is far from optimal. 
    In fact, modern formulations of natural gradient descent—such as NTK or minSR—require inverting a matrix whose size scales quadratically with the number of samples, imposing a practical upper limit of roughly $2^{16}$ samples unless further approximations are employed.
    We demonstrate that the variance of empirical gradient estimates strongly depends on the properties of the distribution from which samples are drawn. Crucially, this variance is an inherent property of the estimator, independent of the specific sampling strategy used. We show that sampling from the Born distribution is typically suboptimal and that the variance can be largely reduced by adaptive importance sampling strategies at no additional cost.

    \paragraph{Importance sampling}
    
    A commonly used method to reduce the variance of Monte Carlo estimators is importance sampling, where samples are drawn from an alternative distribution $q$, typically chosen as a function of the quantity being estimated. 
    In deep learning, importance sampling has been applied to accelerate training by emphasizing the most informative data points \cite{alain_variance_2016, katharopoulos_not_2018, liu_adam_2020}, thereby effectively reducing gradient variance. 
    In variational inference, it has also been used both to reduce variance \cite{ruiz_overdispersed_2016, li_variance_2018} and to facilitate the sampling of energy-based models \cite{grenioux_balanced_2024}. 
    In the VMC literature, a few authors have proposed sampling from $\lvert\psi(x)\rvert$, a simple form of importance sampling, to enhance training performance \cite{chen_efficient_2023,Inui_2021}.
    However, this approach is rarely adopted, remains poorly understood, and lacks a clear theoretical justification.
    Nevertheless, this sampling scheme can be interpreted as a special case of overdispersed importance sampling \cite{ruiz_overdispersed_2016},  where samples are drawn from $|\psi|^\alpha$. 
    For $0 < \alpha < 2$ this distribution has heavier tails than $|\psi|^2$, mitigating sampling issues while introducing only a single new hyper-parameter.

    In this manuscript, we systematically investigate importance sampling as a variance reduction technique for optimizing neural quantum states. 
    Inspired by \cite{ruiz_overdispersed_2016}, we restrict to a one-parameter family of overdispersed sampling distributions and propose an adaptive algorithm to tune this additional parameter for improved gradient estimation.
    The manuscript is structured as follows: 
    in \cref{sec:importance-sampling-vmc}, we review the VMC formalism under arbitrary sampling distributions. In \cref{sec:adaptive-overdispersion}, we present our adaptive strategy for optimizing the sampling distribution. Numerical results are reported for a range of benchmark problems: frustrated spin systems on the square lattice (\cref{sec:results-j1j2}), molecular Hamiltonians (\cref{sec:results-molecules}), and infidelity minimization in neural-projected quantum dynamics (\cref{sec:results-fidelity}).

\section{Importance sampling in variational Monte Carlo}
    \label{sec:importance-sampling-vmc}
    In the following, we consider systems composed of $N$ particles. We denote the Hilbert space of such systems by $\mathcal{H}$ and a generic variational state by $\ket{\psi_\theta}\in\mathcal{H}$, where $\theta \in \mathbb{R}^{N_p}$ is the vector of variational parameters. 
    We work in the computational basis $\{\ket{x}\}$ of $\mathcal{H}$. The wavefunction amplitude for a given basis state $x$ is given by $\psi_\theta(x) = \bra{x}\ket{\psi_\theta}$.     
    
    \subsection{Importance sampling for gradient estimation}    
    The prototypical VMC framework focuses on minimizing objective functions expressed as statistical averages over the Born distribution $p(x) = |\psi_\theta(x)|^2$ with normalization constant $Z_{\theta} = \braket{\psi_\theta} = \sum_x p(x)$. 
    The general loss function takes the form $\mathcal{L}(\theta) = \E_{p}[\ell(x)]$, for some local quantity $\ell(x)$. 
    For example, in ground-state optimization problems $\ell(x) = H_{\mathrm{loc}}(x) = \mel{x}{H}{\psi_{\theta}}/\braket{x}{\psi_\theta}$ is the local energy of some Hamiltonian $H$. 
    In state compression tasks, the same form is used with $H = \ketbra{\phi}/\braket{\phi}$ and for a fixed target state $\ket{\phi}$.

    The optimization of $\mathcal{L}$ is driven to convergence by consecutive estimates of the gradient $\bm F = \grad_{\theta} \mathcal{L}(\theta) \in \mathbb{R}^{N_p}$, which can itself be written as a statistical average: $F_i = \E_{p}[f_i(x)] \in \mathbb{R}$. 
    In standard NQS implementations such as those mentioned above, the local gradient components $f_i(x)$ are given by 
    \begin{equation}
    \label{eqn:local_grad}
        f_i(x) = 2\,\Re{\partial_{\theta_i}\!\log \psi_\theta(x)\,\,\Delta \ell(x)^*},
    \end{equation}
    where $\Delta A(x) \equiv A(x) - \E_{p}[A]$ denotes a centered observable. 
    Practical finite-sample estimates of the true gradient $\bm F$ are obtained by drawing $N_s$ i.i.d.~samples from $p(x)$ and forming the Monte Carlo estimator 
    \begin{equation}
        \hat F_i = \frac{1}{N_s}\sum_{\mu=1}^{N_s} f_i(x_\mu), \qquad x_\mu \sim p
    \end{equation}
    such that $\V[\hat F_i] = \V[F_i]/N_s$.
    To ensure reliable convergence, it is crucial to accurately estimate the direction of steepest descent
    $\bm F$ from the background noise introduced by finite Monte Carlo sampling. 
    Our ability to do so can be quantified in terms of the signal-to-noise ratio (SNR) vector, defined as $\operatorname{SNR}_p(\bm F) = (\operatorname{SNR}_p(f_1), \ldots, \operatorname{SNR}_p(f_{N_p}))$ with 
    \begin{equation}
        \operatorname{SNR}_p(f_i) = \sqrt{\frac{\abs{\E_p[f_i(x)]}^2}{\V_p[f_i(x)]}}.
    \label{eq:SNRcomponents}
    \end{equation}
    Each parameter direction can thus be resolved with varying degrees of statistical accuracy. 
    When $\operatorname{SNR}_p(\hat{F_i}) =\sqrt{N_s}\operatorname{SNR}_p(f_i) < 1$, noise dominates the signal, rendering accurate estimation of the gradient difficult or impossible, thereby hindering reliable progress in the optimization.
    As a normalized, dimensionless quantity, the SNR provides more reliable insight than the raw variance, particularly when dealing with vector estimators as it is insensitive to the scale of each component.
    This notion has already appeared in the NQS literature, mainly in the study of dynamics \cite{schmitt_quantum_2020, sinibaldi_unbiasing_2023, gravina_neural_2024}.

    To improve the resolution with which the gradient is resolved, we use importance sampling: we sample from an alternative distribution $q(x)$ that is more correlated with the quantity of interest, thereby reweighting the estimator under this new distribution.
    Specifically, we employ the identity 
    \begin{equation}
        F_i = \E_{p}[f_i(x)] = \E_{q}[W(x) f_i(x)],
    \end{equation}
    where $q(x)$ is a generic importance sampling distribution with normalization $Z_q = \sum_x q(x)$, and
    \begin{equation}
           W(x) = \frac{|\psi_\theta(x)|^2}{q(x)}\frac{Z_q}{Z_\theta}
           \label{weight}
    \end{equation}
    are the importance weights. When $q(x)=\abs{\psi_\theta(x)}^2$, the weights reduce to $W(x)=1$ and we recover the standard estimator. 
    The finite-sample estimator of $F_i$ under $q(x)$ is 
    \begin{equation}
        \hat F_i = \frac{1}{N_s} \sum_{\mu=1}^{N_s} f_i^q(x_\mu), 
        \qquad f_i^q(x) = W(x) f_i(x),
    \end{equation}
    where $x_\mu \!\sim\! q$.
    When sampling from $q$, any centered quantity is considered centered with respect to the sampling distribution. Specifically,
    \begin{equation}
    \label{eqn:centering}
    \begin{aligned}
        \Delta A(x) = A(x) - \E_p[A(x)] = A(x) - \E_q[W(x)A(x)].
    \end{aligned}
    \end{equation}
    Accordingly, this affects the definition of $f_i$ in \cref{eqn:local_grad}.
    Different choices of $q(x)$ produce estimators with different variance properties and thus different SNR vectors $\operatorname{SNR}_q(\bm F) = (\operatorname{SNR}_q(f_1^q), \ldots, \operatorname{SNR}_q(f_{N_p}^q))$. 
    We evaluate the performance of a given distribution $q$ using the average SNR across all parameter directions:
    \begin{equation}
        \label{eq:snr-loss}
        \mathcal{L}_{\rm IS}(q) 
        = \frac{1}{N_p}\sum_{i=1}^{N_p} \mathrm{SNR}_q\bigl(f_i^q\bigr).
    \end{equation}
    Importantly, $\mathcal{L}_{\rm IS}$ can be expressed as a ratio of expectations under $q$, which in turn can be estimated by Monte Carlo sampling. 
    Specifically, the variance $\V_q[f_i^q(x)]$ appearing at the denominator of Eq.~\eqref{eq:SNRcomponents} can be expressed as
    \begin{equation}
        \V_q[f_i^q(x)] = \E_{q}\left[W^2(x)\,|f_i(x) - F_i|^2\right].
        \label{eq:varis}
    \end{equation}
    A detailed derivation of \cref{eq:varis} is provided in \cref{appendix:snis}.
    A practical consequence of this is that $\mathcal{L}_{\rm IS}$ can be evaluated  during the optimization at no additional cost, provided the full Jacobian has already been materialized.
    This is indeed the case for \textit{vanilla} SR/NTK implementations as those described in Refs.~\cite{chen_empowering_2024,rende_simple_2024}.
    Nonetheless, we remark that the square modulus in \cref{eq:varis} prevents us from applying the layer-wise contraction trick from Ref.~\cite{novak2022ntk} that is known to further speedup the computation of the NTK.

    Finally, we remark that in VMC the normalizing constants $Z_\theta$ and $Z_q$ are not known a priori and the ratio $Z_q/Z_\theta$ appearing in \cref{weight} must itself be empirically estimated. 
    This can be done by expressing $W(x) = w(x) / \E_{q}[w(x)]$ with $w(x) = |\psi_\theta(x)|^2 / q(x)$ and $\E_{q}[w(x)] = Z_\theta/Z_q$. This leads to the so-called \emph{self-normalized} importance sampling estimator which is in general biased. For a detailed discussion, see \cref{appendix:snis}.

    \subsection{Optimal importance sampling distribution}
    \label{sec:optimal-is}
    Since the gradient $\bm F$ is a vector of statistical averages, maximal variance reduction across each component in general requires sampling from $N_p$ distinct distributions, each tailored to the local structure of the corresponding gradient direction.  
     According to Hesterberg \cite{hesterberg_weighted_1995}, the optimal sampling distribution that minimizes the variance of the self-normalized importance sampling estimator of a given gradient component $F_i$ is
    \begin{equation}
        \label{eq:optdistrib}
        q_{\rm opt}^i(x) \propto |\psi(x)|^2\,|f_i(x) - F_i|.
    \end{equation}
    In practice, this strategy is impractical in VMC applications due to its high computational cost. 
    It requires prior knowledge of the exact value of $F_i$, evaluation of derivatives at every Monte Carlo step, and mostly the ability to sample separately from a different distribution for each parameter $i$, which becomes prohibitively expensive in high-dimensional settings. 
    Nevertheless, this distribution remains valuable as a theoretical benchmark, offering insight into how an ideal sampler would allocate probability mass to reduce variance most effectively.

        \begin{figure}[t!]
        \includegraphics[]{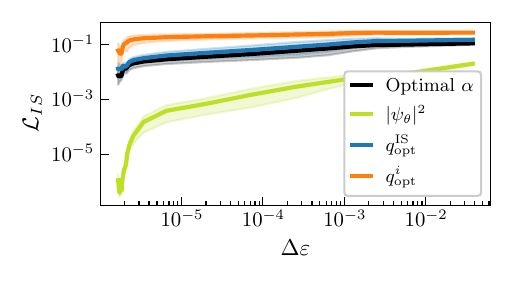}
        \caption{
        Component-wise averaged signal-to-noise ratio (SNR), as defined in \cref{eq:snr-loss}, measured along the ground-state optimization trajectory of the $\mathrm{N}_2$ molecule. The reference trajectory is computed in full summation, i.e., by evaluating all statistical averages exactly by summing over the entire Hilbert space. For each state visited during the optimization, we report the SNR of noisy energy gradient estimates obtained using four sampling strategies. First, the standard Born distribution $|\psi_\theta(x)|^2$ (green). Second, the per-component optimal importance sampling distribution $q{\rm opt}^i$ from \cref{eq:optdistrib} (orange), which requires $N_p$ distinct sampling processes. Third the uniform mixture of optimal importance sampling distributions $q_{\rm opt}^{\rm IS} \propto \sum_i |\psi_\theta(x)|^2\, |f_i(x)|$ (blue). Fourth, the overdispersed parametric distribution $|\psi_\theta(x)|^\alpha$ with $\alpha$ optimized at each optimization step to maximize the SNR (black). The horizontal axis shows the relative error of the variational energy with respect to the exact ground-state energy, rather than the iteration index, to enhance interpretability.
        }
        \label{fig:gradvsnr}
    \end{figure}

    \subsection{Overdispersed distributions}
        \label{sec:overdispersed}
        In this work we consider the family of sampling distributions defined as
        \begin{equation}
            \qty{q_\alpha(x) = \frac{|\psi_\theta(x)|^\alpha}{Z_\alpha},\quad \alpha\geq 0},
        \end{equation}
        where $Z_\alpha = \sum_x q_{\alpha}(x)$ is the normalization constant.
        When $\alpha<2$, these distributions are known as \emph{overdispersed} \cite{ruiz_overdispersed_2016}, as they flatten the tails of the Born distribution and converge to the uniform distribution in the limit $\alpha\to 0$.
        We will show that, in general, there exists an $\alpha$ for which the variance of the gradient estimator in Eq.~\eqref{eq:varis} is lower than that obtained using standard Born sampling ($\alpha = 2$).
        While it may seem counterintuitive to underweight the regions where the wave function amplitude is largest, overdispersed sampling often leads to more accurate gradient estimates.
        There are two key reasons for this improvement.
        First, the optimal variational state $\ket{\psi_{\theta^*}}$ may be highly localized, causing standard sampling to explore only a narrow region of the Hilbert space, potentially too limited to capture meaningful gradient information.
        Second, flatter sampling distributions typically lead to improved mixing times and acceptance rates in MCMC methods, enabling more efficient exploration of the configuration space.

        In \cref{fig:gradvsnr}, we report the performance of various gradient estimators along a prototypical ground state optimization trajectory. The performance is quantified using the component-averaged SNR, $\mathcal{L}_{\rm IS}(q)$, as defined in \cref{eq:snr-loss}.
        We compare four sampling strategies. First, sampling from the standard Born distribution $|\psi_{\theta}(x)|^2$, whose SNR rapidly vanishes as convergence is approached. Second, independent sampling from the $N_p$ optimal distributions $q_{\rm opt}^i$, each tailored to a single gradient component, as described in \cref{sec:optimal-is}. In line with \cite{hesterberg_weighted_1995}, this method achieves the slowest decay in SNR but is prohibitively expensive for large system sizes.
        Third, we consider sampling from a single distribution defined as a mixture of the optimal importance sampling distributions, $q_{\rm opt}^{\rm IS} = \frac{1}{N_p} \sum_i |\psi_\theta(x)|^2\, |f_i(x)|$ \footnote{We do not include a mixture of the exact optimal distributions $q_{\rm opt}^i$ as estimating each $F_i$ at every MCMC step would be computationally prohibitive and largely redundant.}. 
        A detailed discussion of the rationale behind this choice can be found in \cref{appendix:kl}.
        Finally, we examine sampling from the overdispersed distribution $|\psi_\theta(x)|^\alpha$, where the value of $\alpha$ is optimized at each step to maximize the SNR. 
        We find that this approach performs comparably to $q_{\rm opt}^{\rm IS}$, with both achieving nearly the same accuracy as the optimal component-wise sampling. This supports the use of $|\psi_\theta(x)|^\alpha$ as a practical and effective parametric family for importance sampling.
        
        Note that sampling from the overdispersed distribution  introduces no additional computational overhead, but it adds a new hyperparameter $\alpha$ to set, and possibly tune, for each problem.
        The computational cost of VMC optimizations motivates the development of an adaptive tuning strategy for $\alpha$.
      
\subsection{Adaptive overdispersion strategy}
    \label{sec:adaptive-overdispersion}

    Inspired by Ref.~\cite{ruiz_overdispersed_2016}, we restrict our search space to the one‐parameter family of overdispersed distributions $q_\alpha(x)\propto |\psi_\theta(x)|^\alpha$ and adaptively tune $\alpha$ to \textit{maximize} $\mathcal{L}_{\rm IS}$. 
    Rather than relying on costly hyperparameter searches, we update $\alpha$ dynamically at each optimization step using a local gradient-based rule:
     \begin{equation}
        \label{eq:alpha-update-rule}
        \mathbf{\alpha}' = \mathbf{\alpha} + \eta\, \partial_\alpha\mathcal{L}_{\rm IS}(q_{\alpha}),
    \end{equation}
    where $\eta>0$ is a fixed, conservative, learning rate.
    Since the optimal value of $\alpha$ depends on both the Hamiltonian and the current variational state, this approach naturally adjusts to changes in the variational state as the optimization progresses.
    The gradient of the importance sampling objective $\mathcal{L}_{\mathrm{IS}}(q_\alpha)$ with respect to the overdispersion parameter $\alpha$ is found by recursive application of the chain rule: 
    \begin{equation}
    \begin{aligned}
        \partial_\alpha \mathcal{L}_{\mathrm{IS}}(q_\alpha) 
        &= \frac{1}{N_p} \sum_{i=1}^{N_p} \partial_{\alpha} \operatorname{SNR}_{q_\alpha}\bigl(f_{i}^{q_\alpha}\bigr),\\[0.5em]
        \partial_{\alpha} \mathrm{SNR}\qty(f_i^{q_\alpha}) &= 
        \frac{\partial_{\alpha}\log\qty(\V_{q_\alpha}\qty[f_i^{q_\alpha}])}{2 \, \mathrm{SNR}\qty(f_i^{q_\alpha})},
    \end{aligned}
    \end{equation}
    where 
    \begin{equation}
    \begin{aligned}
         &\partial_{\alpha}\! \V_{q_\alpha}[f_i^{q_\alpha}(x)] = -
         \operatorname{Cov}_{q_\alpha}\!\bigl(\,\partial_{\alpha}\log q_\alpha(x)\,,\, g_i(x)^2\,\bigr),\\[0.5em]
         &g_i(x) = W(x)\,|f_i(x) - F_i|.
    \end{aligned}
    \end{equation}

 In practice, we find that a learning rate of $\eta=0.1$  yields stable and effective convergence across all tested cases, enabling robust, fully automated tuning of the overdispersion factor. 
 To preserve Markov chain thermalization—since the sampling distribution evolves during training—we clip the per-step update of $\alpha$ typically restricting it to a maximum increment of $0.01$ so as not to excessively diverge from the previous iterate.

  \begin{figure*}
        \centering
        \hspace*{-1.2em}
        \includegraphics[]{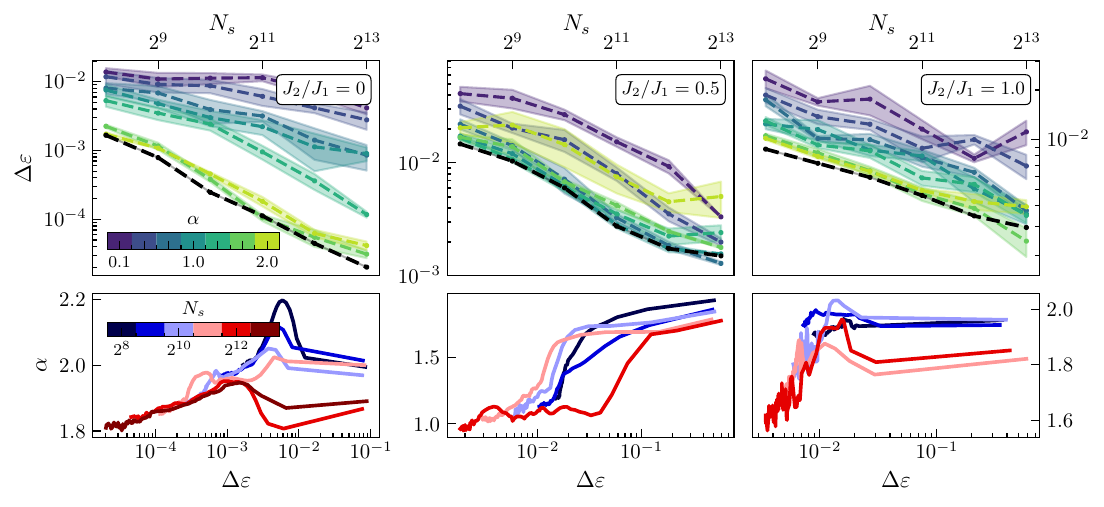}
        \caption{Scaling of relative error with sample size for a square $J_1$-$J_2$ model with $L=6$, for several importance sampling strategies and couplings.
        In both cases, the adaptive strategy (black) outperforms the fixed overdispersion coefficient. (\textbf{Left}): $J_1$-$J_2$, $J_2=0.5$. (\textbf{Middle}): $J_1$-$J_2$, $J_2=1.0$. (\textbf{Right}): Heisenberg model. The adaptive trajectories are shown in the bottom panels. We use $n_l = 4$ layers, $d_{\rm model} = 24$ features and $h = 12$ heads. We impose only translational symmetry, disregarding other symmetries, and, consistent with previous ViT-based approaches, we do not impose a Marshall prior on the basis \cite{marshall_antiferromagnetism_1955}. With this setting the network has $N_p \approx 10^4$ parameters, and is thus optimized using the tangent kernel method. The plotted variational energies are averaged over 10 runs. }
        \label{fig:gsj1j2}
    \end{figure*}
  
    \subsection{Stochastic reconfiguration}
        \label{sec:stochastic-rec}
        Successful ground-state and fidelity optimizations often require stochastic reconfiguration to effectively navigate the rugged optimization landscape characteristic of such problems \cite{sorella_green_1998,Amari1998,chen_empowering_2024, bukov_learning_2021}.
        Stochastic reconfiguration preconditions the bare gradient update $\grad_\theta\mathcal{L}(\theta)$ using the inverse of the Fubini-Study metric tensor, which encodes the local geometry of the Hilbert space at the current parameter configuration. Specifically, the update rule takes the form 
        \begin{equation}
            \theta_{k+1} = \theta_{k} - \alpha_k\,(\bm S + \lambda_k)^{-1}\grad_\theta\mathcal{L}(\theta),
        \end{equation}
        where $\alpha_k$ is the learning rate and $\lambda_k$ is a form of Tikhonov regularization for $\bm S$ \cite{Wright2006}. 
        The components of $\bm S$ are given by
       \begin{equation}
       \label{eqn:qgt}
        \begin{aligned}
            S_{ij} 
            &= \Re{\E_{q}\!\left[W(x) \, \Delta\partial_{\theta_i}\!\log \psi_\theta(x)^*\,\,\Delta\partial_{\theta_j}\!\log \psi_\theta(x)\right]},
        \end{aligned}
        \end{equation}
        where the centering (denoted by $\Delta$) is performed with respect to the sampling distribution $q$, as described in \cref{eqn:centering}.
        For practical implementations, it is important that both the gradient and the metric tensor are computed over the same set of samples \cite{Becca_Sorella_2017}. This ensures that the stochastic errors in both estimators are correlated, leading to reduced variance in the natural gradient update. 
        Accordingly, the sampling distribution $q$ should be shared between the two estimators.

        When targeting SR, one may opt for a sampling distribution that improves the SNR of (i) the quantum geometric tensor, (ii) the variational forces, (iii) or some quantity that depends on both.
        In section 3 of Ref.~\cite{gravina_neural_2024} it was shown that optimal estimators can be identified by focusing solely on the gradient estimator, with the geometric tensor playing only a minor role.
        Furthermore, stable VMC optimizations with geometric tensors computed over as little as $\approx100$ samples have been reported by Malyshev et al.~\cite{malyshev_neural_2024}.
        In line with these findings, in this work we focus exclusively on maximizing the SNR of the bare gradient when comparing different sampling distributions, as discussed in the previous sections. This assumption will be validated a posteriori by the performance of our approach.

\section{Applications and Numerical Results}
    
    In this section, we present a comprehensive numerical study of the effects of overdispersed sampling and our autotuning strategy on the simulation of various physical systems. We begin with spin systems (\cref{sec:results-j1j2}) and molecular Hamiltonians (\cref{sec:results-molecules}), and then extend the analysis to state-matching problems characteristic of p-tVMC-based dynamical simulations (\cref{sec:results-fidelity}).
    
    \subsection{Spin systems : $J_1$–$J_2$ and Heisenberg model}
        \label{sec:results-j1j2}
    
         To evaluate the performances of IS in VMC in a realistic setting, we study the spin-$1/2$  $J_1$-$J_2$ model on the square lattice, a common benchmark for NQS.
         The Hamiltonian is 
        \begin{equation}
            H = J_1\sum_{\langle i,j \rangle} \mathbf{S}_i \cdot \mathbf{S}_j + J_2 \sum_{\langle \langle i,j \rangle \rangle} \mathbf{S}_i \cdot \mathbf{S}_j,
        \end{equation}
        where $\mathbf{S}_i = (S_i^x, S_i^y, S_i^z)$ denotes the spin-$1/2$ operator at site $i=1,...,L^2$, and $\langle...\rangle$, $\langle \langle...\rangle \rangle$ respectively indicate pairs of nearest and next-nearest neighbors sites.
        
         It has been shown that finding the ground state of this model close to the point of maximum frustration $J_2/J_1 = 0.5$ is a hard problem for NQS.
         Bukov et al.~\cite{bukov_learning_2021} conjectured that the optimization landscape in this regime is highly rugged, characterized by numerous local minima.
         As a result, training NQS in such a non-convex landscape is exceptionally difficult.
    
         Accurate gradient estimation is therefore essential for ensuring convergence.
         This can be achieved either by increasing the number of samples or by employing IS techniques.
         In this section, the variational ansatz we use a reasonably-sized, factored-attention Vision Transformer (ViT) \cite{viteritti_transformer_2023}, which has shown strong performances on this system \cite{rende_simple_2024}.
         Architectural details are provided in \cref{appendix:vit}.
        
        Fig.~\ref{fig:gsj1j2} shows how importance sampling improves performance over traditional Born sampling. We study the scaling of the relative error $\Delta \varepsilon$ with the number of samples $N_S$, for several overdispersion coefficient $\alpha \in [0,2]$. All curves have a similar scaling, though not with the same prefactor. While the improvement obtained from IS is minimal for the Heisenberg model ($J_2/J_1 = 0$) or the $J_1$-$J_2$ model at $J_2/J_1 = 1.0$, we observe a substantial decrease in relative error at  $J_2/J_1 = 0.5$, where the improvement is of roughly half an order of magnitude. Moreover, we study the performances of  importance sampling with an auto-tuned overdispersed distribution, depicted by the black curve, which consistently yields the best results across all scenarios. In the bottom row of the plots are plotted the different trajectories of the adapted coefficient against the relative error, showing a consistent behavior across sample sizes.
        
        \begin{figure}[t]
            \includegraphics[]{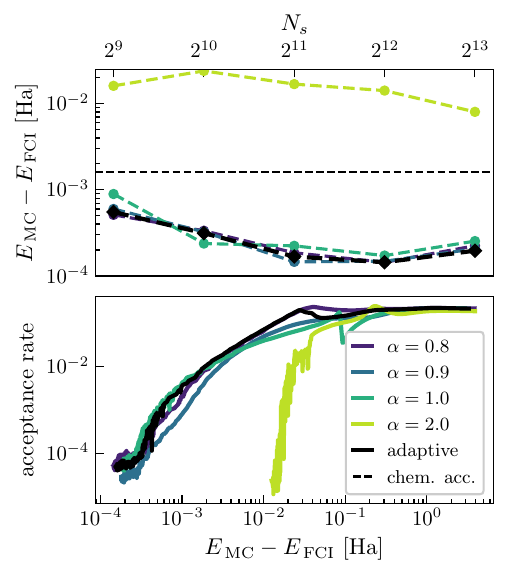}
            \caption{(\textbf{Top}): scaling of the best VMC absolute error with respect to the Full Configuration Interaction (FCI) energy of Lithium Oxide in the STO-3G basis as a function of the number of samples. The dashed line denotes chemical accuracy.
            (\textbf{Bottom}): acceptance rate for several sampling distributions over an optimization performed with $N_S=2^{12}$ samples.}
            \label{fig:li2O}
        \end{figure}
        
        \begin{figure*}
            \includegraphics[]{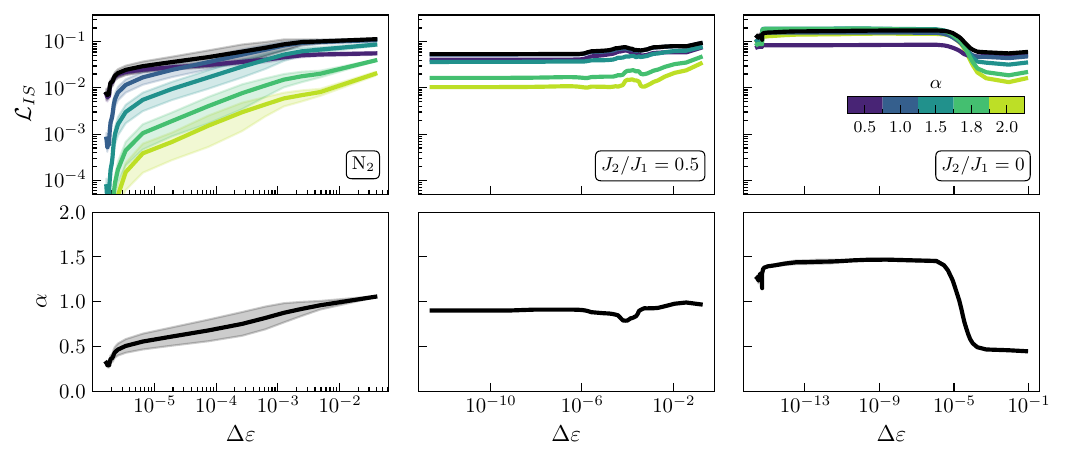}
            \caption{Mean gradient Signal-to-Noise ratio, for several importance sampling distribution and systems. (\textbf{Top}): show the evolution of the SNR as a function of the relative error to the exact ground state. The black curve denotes the maximal SNR one can attain using overdispersed sampling distributions. (\textbf{Bottom}): evolution of the overdispersion coefficient which maximizes the SNR. \textbf{Left}: Nitrogen in the STO-3G basis ; \textbf{Middle}: square $J_1$-$J_2$, $L=4$, $J_2/J_1=0.5$; \textbf{Right}: Heisenberg model, i.e. $J_2=0$. }
            \label{fig:snrcomp}
        \end{figure*}
            
    \subsection{Molecular Hamiltonians in second quantization}
    \label{sec:results-molecules}
    
         The Hamiltonians of molecular systems,
        \begin{equation}
        H=\sum_{i,j,\sigma}t_{ij}c_{i,\sigma}^\dagger c_{j,\sigma}+\frac{1}{2}\sum_{i,j,k,l,\sigma,\sigma^{\prime}}V_{ijkl}c_{i,\sigma}^\dagger c_{j,\sigma^{\prime}}^\dagger c_{l,\sigma^{\prime}}c_{k,\sigma},
        \end{equation}
        where $t_{ij}$ and $V_{ijkl}$ depend on the exact choice of molecule and basis, in general lead to very peaked wave-functions, which makes them challenging for VMC-based approaches to tackle \cite{choo_fermionic_2020}.
        Alternatives like autoregressive models \cite{malyshev_autoregressive_2023, malyshev_neural_2024} have been proposed to balance for the inefficiency of the MCMC sampling.
        Sophisticated sampling without replacement techniques like the Gumbel top-k trick \cite{kool_stochastic_2019} has been applied to enlarge the exploration of the Hilbert space, either on the whole space with autoregressive models \cite{malyshev_autoregressive_2023} or on an evolving subset of configurations \cite{liu_efficient_2025}.
        
        In this section, we investigate the impact of importance sampling on the ground-state search for the $\rm Li_2O$ molecule whose geometry is retrieved from the PubChem \cite{noauthor_pubchem_nodate} database, using the STO-3G basis and canonical Hartree-Fock (HF) orbitals; in that setting, $\mathrm{dim}(\mathcal{H}) = 4.14 \times 10^7$.
        We use a neural network backflow with parameters identical to the literature \cite{liu_neural_2024, liu_efficient_2025}.
        While it has been identified as a particularly challenging molecule for which standard MCMC struggles, we are able to improve the results by at least two orders of magnitude using automatic overdispersed sampling, surpassing chemical accuracy.
        In the top panel of \cref{fig:li2O}, we plot the scaling of the absolute energy error with respect to the Full Configuration Interaction (FCI) energy of Lithium Oxide in the STO-3G basis as a function of the number of samples. We are able to approximately reach the same accuracy as the one obtained in Ref.~\cite{liu_neural_2024} with fixed-size selected configuration (FSSC) scheme and $2^{14}$ samples, using as little as $2^{11}$ samples, while the error of traditional Born sampling remains two orders of magnitude bigger. 
        It is worth noting that the optimization remains particularly challenging and is also limited by the low acceptance rate of MCMC chains, which approaches $\mathcal{O}(10^{-4})$ near convergence, as depicted in the bottom panel of \cref{fig:li2O}, where one can see that $|\psi|^2$ struggles the same way even though further from the ground state.
        
        This could be improved by designing a transition rule more fitted to the Hamiltonian, which introduces long-range correlations.
        Currently, the sampler proposes new configuration by exchanging two orbitals of the same spin along the edges of the graph defined by the adjacency matrix $A_{ij} = t_{ij} + \sum_k V_{kikj}$.
        
        We found that using the SPRING optimizer \cite{goldshlager_kaczmarz-inspired_2024} with $\mu = 0.9$ improved the overall stability; we refer the reader to \cref{appendix:hparams} for more details on the experimental setup.
        
           
    \subsection{Signal-to-Noise Ratio analysis}
        \label{sec:results-snr}
         To isolate the intrinsic benefits of importance sampling from Monte Carlo noise, in \cref{fig:snrcomp} we perform exact evaluations over the full Hilbert space, summing over all basis states without sampling.
         This allows us to compute $\mathcal{L}_{\rm IS}$ exactly and determine the optimal overdispersion parameter at various stages of training.
    
         We find that the effect of importance sampling on the SNR depends strongly on the target system.
         In the $J_1–J_2$ model, in the case of a $4\times 4$ system, significant SNR gains appear only near maximal frustration ($J_2/J_1=0.5$), where the Born distribution poorly captures critical configurations.
         By contrast, we examine the nitrogen molecule $N_2$ in the STO-3G basis ($\dim(\mathcal{H}) = 1.44\times10^4$), a classic strongly correlated system with a highly peaked ground state.
         In this regime, the extremely small values of $|\psi_\theta(x)|^2$ impedes the estimation of the gradient because some configurations with a high gradient magnitude will almost never be sampled.
         Our analysis confirms this effect: near convergence, $\mathcal{L}_{\rm IS}(|\psi_\theta|^2)$ vanishes.
         Moreover, we observe at most a two-orders-of-magnitude difference between the SNR of $|\psi_\theta(x)|^2$ and the optimal \emph{overdispersed} distribution.
         This difference implies that the number of required samples for a Monte-Carlo estimator under $|\psi_\theta(x)|^2$ to achieve a comparable SNR to the optimal one must be increased by at least a factor of 1000.
    
        A pratical stability criterion, proposed in Sinibaldi et al., \cite{sinibaldi_unbiasing_2023}, states that a Monte Carlo estimator of the gradient with $N_{s}$ samples becomes reliable once $\sqrt{N_{s}}\mathcal{L}_{\rm IS}(q) \gtrsim 1$.
        This could provide a criterion for adaptive sample size;
        Indeed, during the VMC optimization one can dynamically adjust the sample size to maintain $\sqrt{N_{s}}\mathcal{L}_{\rm IS}(q) \gtrsim 1$.
        
        Finally, it is important to mention that because of ergodicity, VMC can still perform decently even though the SNR of the gradient is low. However, a higher SNR systematically yields greater stability and faster convergence in practice.
    
         \begin{figure}[t]
            \center
            \includegraphics[]{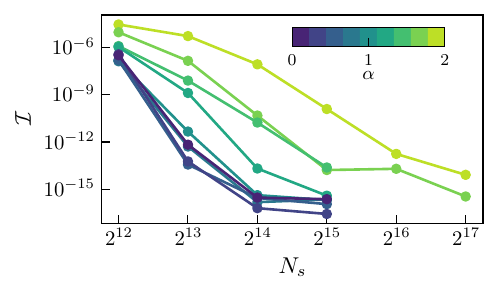}
            \caption{Effect of varying the overdispersion parameter $\alpha$ on a prototypical Infidelity Optimization for variational dynamics taken from Ref. \cite{gravina_neural_2024}. The curve shows the exact final infidelity at convergence after 500 natural gradient optimization steps. The hyperparameters are provided in \cref{appendix:infid}.}
            \label{fig:infidelity}
        \end{figure}    
            
    \subsection{Infidelity optimization for p-tVMC}
    \label{sec:results-fidelity}
    
        We complete our numerical tests by applying the same framework to an \textit{infidelity minimization}, sometimes referred to as \textit{state compression} task.
        This kind of optimization is central to projected quantum dynamics~\cite{sinibaldi_unbiasing_2023,gravina_neural_2024,Nys24fermionsdynamics} but could also be used to find excited states or build other algorithms.
    
        We consider a prototypical challenging case of an entangled state evolving during a quench, where the fidelity between the variational state $\ket{\psi_{\theta_{t+dt}}}$ and the time evolved state $e^{-idtH}\ket{\psi_{\theta_t}}$ is maximized.
        Even with refined estimators, it has been observed that solving infidelity optimization problems to machine precision required a very large sample size, much larger than the dimension of the underlying Hilbert space (see Appendix F of Ref.~ \cite{gravina_neural_2024}).
        Importance sampling can be applied in a straightforward way to infidelity optimization, for which the technical details can be found in Appendix \ref{appendix:infid}.
       
        In \cref{fig:infidelity}, we present the state optimization for the update at time $Jt = 1.25 \rightarrow 1.30$ when performing a full-basis simulation of the quench dynamics of the 2D transverse field Ising Model
        \begin{equation}
            H = -4J\sum_{\langle i,j\rangle } S^z_iS^z_j - 2h\sum_iS^x_i
        \end{equation}
        from $h = \infty$ to $h = h_c/10$, on a $4\times 4$ square lattice.
        Reproducing the exact summation result with a Monte Carlo simulation is much easier when using importance sampling, where it is reached using $2^{13}$ samples.
        Conversely, sampling from $|\psi_\theta|^2$ required at least $2^{17}$ samples to reach the same level of precision, a number that even exceeds the Hilbert-space dimension of $2^{16}$ for this system.
        This opens the perspective of a significant speedup and increased reliability of  projected time-dependent VMC (p-tVMC) methods.
        
\section{Conclusion}
    
    In this work, we have shown that importance sampling offers a simple yet powerful lever to reduce gradient noise and accelerate Variational Monte Carlo (VMC) training of neural quantum states.
    We have shown that even an extremely simple one-parameter family of overdispersed sampling distributions, $q_\alpha(x)\propto|\psi_\theta(x)|^\alpha$ can lead to considerable improvements.
    Moreover, we have introduced a technique to optimally tune the single hyperparameter $\alpha$ at a negligible cost when performing natural-gradient optimization.
    
    Across several realistic problems, such as the ground-state search in the frustrated $J_1$–$J_2$ Heisenberg model, the ${\rm Li_2O}$ and ${\rm N_2}$ molecules, and infidelity minimization in projected dynamics, we observe consistent, sometimes dramatic, reductions in required sample counts.
    We remark that at the beginning of this research project we also considered more complex and costly importance sampling distributions incorporating information about the Hamiltonian or the model's Jacobian, but found that they ultimately underperformed or were only marginally better when compared against the much simpler and cheaper overdispersed family.
    
    Our procedure does not increase the computational complexity of Stochastic Reconfiguration yet yields large stability and convergence benefits.
    We believe it can be readily adopted in a wide range of contexts including impurity problems, open quantum systems, strongly correlated fermionic models, and non-equilibrium dynamics.

    As a future research direction, we believe it might be interesting to investigate the use of a generative model as an IS distribution, dynamically optimized to keep the SNR minimal while ensuring perfectly uncorrelated samples.
    Finally, we believe that the success of this approach based on the SNR suggests that more about VMC convergence could be understood by looking at this quantity.
    
    The methods presented in this paper are fairly straightforward to add to an existing VMC codebase. 
    The equations necessary to obtain an implementation are provided in the recapitulative \cref{app:eq-summary-impl-aid}.
    An implementation of the adaptive sampling algorithms discussed in this manuscript, built on top of NetKet \cite{vicentini_netket_2022}, as well as all accompanying data can be found at the repository \href{https://github.com/NeuralQXLab/importance_sampling_nqs}{GitHub@NeuralQXLab/importance\_sampling\_nqs}. 
    
\section*{Acknowledgments}

    Simulations were performed with NetKet~\cite{carleo_netket_2019, vicentini_netket_2022}, and at times parallelized with mpi4JAX~\cite{hafner_mpi4jax_2021}.
    Infidelity optimizations where performed with the code from p-tVMC~\cite{gravina_neural_2024}.
    This software is built on top of JAX \cite{jax2018github} and Flax \cite{flax2020github}.

    We acknowledge Ahmedeo Shokry for providing the initial importance sampling code, as well as the Neural Network backflow code.
    We acknowledge insightful discussions with Marylou Gabrié and Adrien Kahn, and Markus Schmitt.
    F.V. acknowledges support by the French Agence Nationale de la Recherche through the NDQM project ANR-23-CE30-0018.
    This project was provided with computing HPC and storage resources by GENCI at IDRIS thanks to the grant 2024-A0170515698 on the supercomputer Jean Zay's V100 partition.
    We used approximately 5k V100 hours and 1.5k A100 hours.

\bibliography{bibliography}

\appendix
\section{Methods}    
    \subsection{Variational Monte Carlo}

        \par As explained in the main text, Variational Monte Carlo aims at minimizing the expectation value of the hamiltonian under a variational state $\ket{\psi_\theta}$: 
        \begin{equation*}
            E_\theta =  \frac{\bra{\psi_\theta} H \ket{\psi_\theta}}{\bra{\psi_\theta} \ket{\psi_\theta}}.
        \end{equation*}
        \par By inserting the sum of projector along each element of the orthogonal basis $\{\ket{x}\}$, namely $\mathbb{I} = \sum_x \ket{x}\bra{x}$ one can rewrite the previous expression as an expectation value under the Born amplitude $p(x) = \frac{|\psi_\theta(x)|^2}{\bra{\psi_\theta}\ket{\psi_\theta}}$:

        \begin{equation*}
            E_\theta = \E_{p}\left[ H_{loc}(x)\right],
        \end{equation*}
        
        where: \begin{equation*}
            H_{loc}(x) = \frac{1}{\psi_\theta(x)} \sum_{y, H_{xy} \neq 0} H_{xy} \psi_\theta(y)
        \end{equation*}
        is called the local energy. This quantity is indeed local as it can be evaluated with the knowledge of the NQS $\psi_\theta(\cdot)$ and $\ket{x}$.

\subsection{Infidelity optimization}
\label{appendix:infid}
        
    The fidelity between two pure quantum states is defined as:
    \begin{equation*}
        \mathcal{F}(\ket{\psi}, \ket{\phi}) = 
        \frac{\bra{\psi} \ket{\phi}\bra{\phi} \ket{\psi}}{
        \braket{\psi} \braket{\phi}}, 
    \end{equation*}
    where $\ket{\psi}$ will be identified to the variational state, and $\ket{\phi}$ to the target state.
    The goal will be to minimize the infidelity $\mathcal{I} = 1 - \mathcal{F}$.
    Following \cite{gravina_neural_2024}, we write $\mathcal{F}$ as an expectation value marginalized over the target state
    
    \begin{equation*}
        \mathcal{F}(\ket{\psi}, \ket{\phi}) = \E_{|\psi|^2}[\underbrace{\E_{y \sim |\phi|^2}[F(x,y) \mid x]}_{f(x)}].
    \end{equation*}
    In this form, a Monte Carlo estimator of the fidelity suffers from a vanishing SNR when $\psi$ approaches $\phi$.
    It is convenient to use the control-variate-enhanced estimator to get rid of this problem:
    \begin{align*}
        f(x) &= \E_{y \sim |\phi|^2}[F(x, y) \mid x] \\ \notag
         &= \Re\{A_x(x)\E_{y \sim|\phi|^2}[A_y(y)]\} \;+ \\ \notag &c \left( \left| A_x(x) \right|^2 \mathbb{E}_{y \sim \pi_\phi} \left[ \left| A_y(y) \right|^2 \right] - 1 \right),
    \end{align*}
    with $A_x(x) = \phi(x)/\psi(x)$ and $A_y(y) = \psi(y)/\phi(y)$.
    This expression leverages the fact that $\E_{x,y\sim |\psi(x)|^2|\phi(y)|^2}[|A_x(x)A_y(y)|^2] = 1$ to reduce the variance.
    Importance sampling is then applied by independently sampling and reweighting by $|\psi(x)|^\alpha$ and $|\phi(y)|^\alpha$.
    We use the same $\alpha$ for both expectation values as $\ket{\psi}$ is expected to resemble $\ket{\phi}$ near convergence.

    \par It is worth noting that projected quantum dynamics already employed importance sampling to bypass the fact that evaluating $\mathcal{F}(\hat{V}\ket{\psi}, \hat{U}\ket{\phi})$ introduces an expensive sampling procedure from $|\bra{x}\hat{V}\ket{\psi}|^2$ and $|\bra{y}\hat{U}\ket{\phi}|^2$ and that our approach builds on top of this reweighting: instead of reweighting with $|\psi(x)|^2$ (resp. $|\phi(y)|^2$), we reweight with $|\psi(x)|^\alpha$ (resp. $|\phi(y)|^\alpha$).
    
\subsection{The self-normalized Importance Sampling estimator}
    \label{appendix:snis}
     In this section, we review the mathematical aspects of the self-normalized Importance Sampling estimator (SNIS). We start by defining the SNIS estimator, then computes its variance and bias. Afterwards, we derive the derivative of the variance of the SNIS estimator for a variational sampling distribution.

\subsubsection{Definition}
    The definition of the Self-normalized importance sampling estimator is given by \cref{eqn:estimator_2} below. 
    To define it, and discuss its properties we start by introducing some notation. 
    \begin{itemize}
        \item $P(x) = p(x)/Z_p$ is the original (or target) Born distribution we want moments under. 
        \item $Q(x) = q(x)/Z_q$ is the proposal distribution from which we actually intend to sample. 
        \item $x_i \sim Q$ for $i=1,\ldots,n$ are the samples. 
        \item $X$ and $Y$ are arbitrary functions of interest and $X_i = X(x_i)$ and $Y_i=Y(x_i)$. 
        \item $w_i= w(x_i) = p(x_i)/q(x_i)$ are the raw importance weights and $W_1 = \sum_i w_i$, $W_2 = \sum_i w_i^2$ are the first and second weight–moments. 
        \item $\tilde w_i = w_i/W_1$ are the self‑normalized weights. 
    \end{itemize}
    We note for future reference that since 
    \begin{equation}
        \E_{Q}[w(x)] = \sum_{x} \frac{q(x)}{Z_q} \frac{p(x)}{q(x)} = \frac{\sum_x p(x)}{Z_q} = \frac{Z_p}{Z_q},
    \end{equation}
    the expectation value under $P$ of some local quantity f can be written as a ratio,
    \begin{align*}
        \E_{P}[f(x)] 
        &= \sum_x \frac{p(x)}{Z_p} f(x) 
        = \sum_x \frac{p(x)}{Z_p} \frac{q(x)/Z_q}{q(x)/Z_q} f(x) \\
        &= \frac{\sum_x \frac{q(x)}{Z_q} w(x) f(x)}{Z_p/Z_q}
        = \frac{\E_Q[w(x)f(x)]}{\E_Q[w(x)]}.
    \end{align*}

    We now define the self-normalized importance sampling estimator as the plug-in estimator of the above ratio, that is 
    \begin{align}
    \label{eqn:estimator_2}
        \E_P[f(x)] 
        &= \notag \frac{\E_Q[w(x)f(x)]}{\E_Q[w(x)]} 
        \approx \frac{\frac{1}{n}\sum_{i=1}^nw_if(x_i)}{\frac{1}{n}\sum_{i=1}^nw_i} \\
        &= \frac{\sum_{i=1}^nw_if(x_i)}{W_1} 
        = \sum_{i=1}^n \tilde w_if(x_i).
    \end{align}

    Let $X$ and $Y$ bet two random variables. Lets introduce some auxiliary notation. Let $\mu_X=\E_P[X]$, $\mu_Y=\E_P[Y]$, $\mu_{XY}=\E_P[XY]$, and $C_{XY}=\cov_P(X,Y) = \E_P[\underbrace{(X-\mu_X)(Y-\mu_Y)}_{C(X.Y)}]$.  Equation~\eqref{eqn:estimator_2} can be applied to the means of $X(x)$ and $Y(x)$ for which we have that
    \begin{equation*}
        \mu_X = \E_P[X] \approx\widehat\mu_X  = \sum_{i=1}^{n}\tilde w_i X_i,
    \end{equation*}
    and similarly for $\mu_Y$. For the expression of the covariance of $X$ and $Y$, we have an analogous replacement:
    \begin{align*}
        C_{XY} 
        &= \E_{P}\bigl[(X(x)-\mu_X)\,(Y(x)-\mu_Y)\bigr]  \\
        &= \frac{\E_{Q}\bigl[w(x)\,(X(x)-\mu_X)\,(Y(x)-\mu_Y)\bigr]}{\E_Q[w(x)]},
    \end{align*}
    with the plug-in estimator being 
    \begin{equation}
    \label{eqn:is_simple_cov}
        \widehat{C}_{XY}^{\rm{\,simple}} =\sum_{i=1}^{n}\tilde w_i
            (X_i-\widehat\mu_X)(Y_i-\widehat\mu_Y)
          =\sum_{i=1}^{n}\tilde w_i X_i Y_i
            -\widehat\mu_X\widehat\mu_Y .
    \end{equation}
    Recall that now $x_i$ are sampled from $Q$, not from $P$.
    We now want to estimate the variance of this estimator. A finite‑sample closed form is algebraically complicated because of the normalizing denominator. Following the overwhelming majority of the literature, we rely on the first-order multi-variate delta method \cite{oehlert_delta_1992} to obtain the asymptotic variance of both estimators.
    
    \subsubsection{Variance of the SNIS sample covariance}

    In this appendix we will prove that the variance of the covariance estimator (\cref{eqn:is_simple_cov}) of the Self Normalized Importance Sampling (SNIS) is
    \begin{equation}
        \label{eq:varformula}
        \V(\hat{F_i^q}) = \frac{1}{N_S}\frac{\E_q\left[w^2(x)|f_i(x) -F_i|^2 \right]}{(\E_q\left[w(x)\right])^2}.
    \end{equation}

    The proof follows.
    We define the random vectors
    \begin{equation*}
        \bm V_i=
        \frac{Z_q}{Z_p}
        \begin{pmatrix}
          w_i\\
          w_i X_i\\
          w_i Y_i\\
          w_i X_i Y_i
         \end{pmatrix},
    \qquad
    \bm \Sigma_n=\frac{1}{n}\sum_{i=1}^{n} \bm V_i
        =\begin{pmatrix}
          A\\ B \\ C \\ D\\
          \end{pmatrix}
    \end{equation*}
    with 
    \begin{align*}
        A &= \notag \frac1n\sum_iw_i,\quad\quad &&B= \frac1n\sum_iw_i X_i,\\ C&=\frac1n\sum_iw_i Y_i,\quad\quad &&D=\frac1n\sum_iw_i X_i Y_i.
    \end{align*}
    Note that the factor $Z_q/Z_p$ in the definition of $\bm V_i$ is deterministic and only serves the purpose of normalizing the expectation values.
    Indeed, the expectations under $Q$ are
    \begin{equation}
    \label{eqn:averages}
    \begin{aligned}
        \bm m \equiv \E_Q[\bm V] = \bigl(1,\;\mu_X,\;\mu_Y,\;\mu_{XY}\bigr)^{T} = \E_Q[\bm \Sigma_n]
    \end{aligned}
    \end{equation}
    where $\bm V$ is a representative of the sample statistics $\bm V_i$. 
   
    Replacing into Eq.~\eqref{eqn:is_simple_cov} we obtain
    \begin{equation*}
        \widehat{C}_{XY}^{\rm{\,simple}} = g(\bm \Sigma_n) = g(A,B,C,D) = \frac{D}{A}-\frac{BC}{A^{2}}.
    \end{equation*}
    To apply the delta method \cite{oehlert_delta_1992} we need to compute the gradient of $g(\bm \Sigma_n)$ and evaluate it at $\bm m$:
    \begin{equation*}
        \begin{aligned}
    \frac{\partial g}{\partial A}
       & =-\frac{D}{A^{2}}+\frac{2BC}{A^{3}}
         \;\;\xrightarrow[]{\eqref{eqn:averages}}\;\;
         -C_{XY} + \mu_X\mu_Y,
         \\[4pt]
    \frac{\partial g}{\partial B}
       & =-\frac{C}{A^{2}}
         \;\;\xrightarrow[]{\eqref{eqn:averages}}\;\;
         -\mu_Y\\[4pt]
    \frac{\partial g}{\partial C}
       & =-\frac{B}{A^{2}}
         \;\;\xrightarrow[]{\eqref{eqn:averages}}\;\;
         -\mu_X\\[4pt]
    \frac{\partial g}{\partial D}
       & =\frac1A
         \;\;\xrightarrow[]{\eqref{eqn:averages}}\;\;
         1,
    \end{aligned}
    \end{equation*}
    so that 
    \begin{equation*}
        \grad g(A,B,C,D)\,\bigl|_{\bm m}= 
    \bigl(
    -C_{XY} + \mu_X\mu_Y,\;
    -\mu_Y,\;
    -\mu_X,\;
    1
    \bigr)^T.
    \end{equation*}
    To conclude, we need the covariance matrix of one draw, that is
    \begin{align*}
        \bm \Sigma &=\cov_q[\bm V] = \E_q[\bm V \bm V^T] - \bm m\bm m^T
    \end{align*}
    
    From the delta method \cite{oehlert_delta_1992}, we have that to leading order in $n$
    \begin{equation}
    \label{eqn:variance_is_simple}
        \V\bigl[\widehat{C}_{XY}^{\rm{\,simple}}\bigr]
          \approx \frac{1}{n}\grad g(\bm m)^{T}\bm\Sigma\,\grad g(\bm m) + o(n^{-1}).
    \end{equation}
    From here on out we will neglect the terms vanishing faster than $1/n$. 
    Computing the leading term by explicitly carrying out the contraction is straightforward but can be quite tedious. Instead we realize that 
    \begin{align*}
        \grad g(\bm m)^{T}\bm\Sigma\,\grad g(\bm m)
        &= \grad g(\bm m)^{T}\cov_Q[\bm V]\,\grad g(\bm m)\\ &= 
        \V_Q[\grad g(\bm m)^T\,\bm V],
    \end{align*}
    where we used that $\grad g(\bm m)^T\,\bm V$ is a scalar quantity to replace the covariance with the variance. Since $\grad g(\bm m)^T$ is deterministic, we have that $\E_Q(\grad g(\bm m)^T\,\bm V) = \bm m$. 
    Now
    
    \begin{flalign*}
        &\grad g(\bm m)^T(\bm V - \bm m) = 
        \notag \frac{Z_q}{Z_p}\cdot\underbrace{\,w\bigl(XY - \mu_YX - \mu_XY - C_{XY}\bigr)\,}_{=\,(X-\mu_X)(Y-\mu_Y) - C_{XY} = C(X,Y) - C_{XY}} \, \\ &+ \, \underbrace{\,C_{XY} - \mu_X\mu_Y + 2\mu_X\mu_Y - \mu_{XY}\,}_{=\;0}\\
        &=  \frac{Z_q}{Z_p} w \bigl( C(X,Y) - C_{XY} \bigr) 
         \equiv \frac{Z_q}{Z_p} w\,H(X,Y), 
    \end{flalign*}
    where 
    \begin{equation*}
        H(X,Y) \equiv C(X,Y) - C_{XY} \equiv (X-\mu_X)(Y-\mu_Y) - C_{XY}.
    \end{equation*}
    Note that $H(X,Y)$ is a functional and should be seen as $H(X,Y)(x) = H(X(x),Y(x))$.
    Finally we have that 
    \begin{equation*}
    \label{eqn:variance_covariance}
    \begin{aligned}
        &\V\bigl[\widehat{C}_{XY}^{\rm{\,simple}}\bigr]
          = \frac{1}{n}\grad g(\bm m)^{T}\bm\Sigma\,\grad g(\bm m) \\
          &= \frac{1}{n} \V_Q\qty[\frac{Z_q}{Z_p}w\,H(X,Y)] 
          = \frac{1}{n} \frac{\V_Q[w\,H(X,Y)]}{(Z_p/Z_q)^2} \\
          &= \frac{1}{n} \frac{\E_Q[w^2\,H^2(X,Y)]}{(\E_Q[w])^2},
    \end{aligned}
    \end{equation*}
    where in the last equality we used that $wH$ is centered in $Q$.
    The plug-in estimator of this quantity is 
    \begin{equation*}
    \begin{aligned}
        &\widehat\V\bigl[\widehat{C}_{XY}^{\rm{\,simple}}\bigr]
        =  \frac{1}{n} \frac{\frac{1}{n}\sum_i w_i^2\,H^2(X_i,Y_i)}{\qty(\frac{1}{n}\sum_i w_i)^2} \\
        &= \sum_i \qty(\frac{w_i}{\sum_i w_i})^2\,H^2(X_i,Y_i) 
        = \sum_i \tilde w_i^2\,H^2(X_i,Y_i) \\[0.1cm]
        &= \sum_i \tilde w_i^2 \bigl[(X_i - \widehat\mu_X)(Y_i - \widehat\mu_Y) - \widehat C_{XY}^{\,\rm{simple}}  \bigr]^2
    \end{aligned}
    \end{equation*}
    
    Going back to the notation of the paper, we see that we get \cref{eq:varformula}.

\subsubsection{Bias}
    In this appendix we prove that the bias of the SNIS estimator is

    \begin{equation}
        \mu_X \left[\frac{1}{n}\left(\underbrace{\frac{\E_Q[w^2(\mu_X - X)]}{\mu_X \E_Q[w]^2}}_{\rho_0}\right)\right].
    \end{equation} 
    The proof follows.
    To simplify the calculation, we will make an approximation by neglecting the covariance structure of the estimator, and only the consider the bias related to the ratio structure. 
    This approximation is motivated by the fact that the covariance will contribute higher order terms which we assume are not dominant here (we also remark that without this approximation it would be very hard to get interpretable analytical results).

    We recall that the SNIS estimator is denoted as $F = \frac{1}{n}\sum_{i=1}^n \tilde{w}_i X_i$. 
    Similarly to the previous section, we will use the delta method to compute the bias. In that case, we need to go to second order as $\E_Q(\frac{Z_q}{Z_p}(w X) - \mu_X) = \E_Q[\frac{Z_q}{Z_p}w - 1] = 0$. 

    Defining:
    \begin{equation*}
    \label{eqn:delta_method_vector_bias}
        \bm V_i=  \frac{Z_q}{Z_p}    
        \begin{pmatrix}
          w_i\\
          w_i X_i\\
         \end{pmatrix},
    \qquad
    \bm \Sigma_n=\frac{1}{n}\sum_{i=1}^{n} \bm V_i
        =\begin{pmatrix}
          A\\ B 
          \end{pmatrix},
    \end{equation*}
    one observes that given $g(A,B) = \frac{B}{A}$ and $\bm m = (\mu_X, 1)^T$, we have:
    \begin{equation*}
        \notag
        \bm \nabla^2 g(\bm m) = 
        \begin{pmatrix}
            \;2\mu_X
            & \;-1\; 
             \\[6pt]
            \cdot 
            & \;0\; 
    \end{pmatrix}.
    \end{equation*}
    Let $r = \frac{Z_p}{Z_q}$. One can thus write:
    \begin{equation*}
           \begin{aligned}
            &\E(F) \approx \mu_X + \frac12\left[ 2\mu_X \frac{1}{n}\V_Q(\frac{w}{r}) - \frac{2}{n}\mathrm{Cov}_Q(\frac{w}{r},\frac{wX}{r})\right] \\
            &= \mu_X \left[1 + \frac{1}{n}\left(\underbrace{\frac{\E_Q[w^2(\mu_X - X)]}{\mu_X r^2}}_{\rho_0}\right)\right],
        \end{aligned} 
    \end{equation*}

    where we denoted as $\rho_0$ the shrinking factor, independent from the sample size $n$.
    One has especially:
    \begin{align*}
        \frac{\V_Q(w)}{r^2} &= \frac{1}{\rm ESS} - 1,
    \end{align*}
    where the effective sample size (ESS), defined as $\frac{\E_Q\left[w(x)\right]^2}{\E_Q\left[w(x)^2\right]}$, is a common metric to measure the similarity between two probability distributions.
    
    It is often employed in the usual context of importance sampling where one tries to approximate an intractable distribution as a mean to sample from it.
    It varies from 0 for distributions with disjoint support, to 1 for identical distributions.
    
    As a consequence, one can infer that the more dissimilar $q$ and $p$ are, the greater the bias will be.
    
\begin{figure}
        \center
        \includegraphics[]{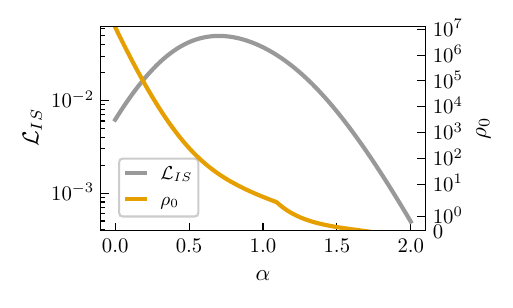}
        \caption{Effective sample size and SNR curves over $\alpha$ computed at the end of a ground state search of the $N_2$ molecule, one of the cases where the optimal $\alpha$ drops to almost $0.5$. The shrinkage factor is relatively small compared to the usually used sample sizes of $N_S \sim O(10^3-10^4)$}
        \label{fig:snr-bias}
    \end{figure}

    In \cref{fig:snr-bias} we plot the signal-to-noise ratio as well as the shrinkage factor $\rho_0$ defined above for a variational state taken at the end of the ground state search for the nitrogen, as described in \cref{sec:results-snr}. One sees that in that case, $\rho_0 \sim O(10^1)$ which is negligible compared to the $\frac{1}{n}$ prefactor as $n$ is usually taken in the range $10^3-10^4$. However, $\rho_0$ quickly increases as $\alpha$ approaches 0, which suggests that using a distribution that is too different from the original sampling distribution may lead to problems, except if one find a way to mitigate this bias by employing a refined estimator.

\subsubsection{Gradient of the SNR}
    \label{appendix:autotuning}
    \par Let $(q_\alpha)_{\alpha \in \mathbb{R}^m}$  be a variational family in $ \mathcal{F}(\{0,1\}^N, \mathbb{R}^+)$.
    Let  $Q_\alpha(x) = \frac{q_\alpha(x)}{Z_\alpha}$ the associated normalized probability distribution.
    Differentiating $Q_\alpha$ with respect to $\alpha$ gives:
    \begin{equation*}
        \forall i \in \{1...m\},
        \frac{\partial_{\alpha_i} q_\alpha(x)}{Z_\alpha} - \sum_y \partial_{\alpha_i} q_\alpha(y) \frac{Q_\alpha(x)}{Z_\alpha},
    \end{equation*}
    which can in general be simplified as: 
    \begin{align*}
        \partial_{\alpha_i} Q_\alpha(x) &= Q_\alpha(x)(\partial_\alpha \log q_\alpha(x) - \E_{q_\alpha}[\partial_{\alpha_i} \log q_\alpha(x)]) \notag \\ &= q_\alpha(x) \Delta_{q_\alpha} \partial_\alpha \log q_\alpha(x),
    \end{align*}
    where 
    \begin{equation*}
        \Delta_{q_\alpha} \partial_\alpha \log q_\alpha(x) = \partial_\alpha \log q_\alpha(x) - \E_{q_\alpha}[\partial_\alpha \log q_\alpha(x)].
    \end{equation*}
    Thus one can write, taking the notations of the main text:
    \begin{align*}
        \partial_{\alpha_i} \V[\hat{F_i}] &= \partial_{\alpha_i} \left( \sum_x \frac{P(x)^2}{Q_\alpha(x)}|f_i(x) - F_i|^2 \right) \notag \\ \notag &=  - \sum_x Q_\alpha(x) \Delta_q\partial_{\alpha_i}\log q_\alpha(x) g_i^2(x)  \\ &= - \E_{q_\alpha}[\Delta_{q_\alpha}\partial_{\alpha_i}\log q_\alpha(x) g_i^2(x)],
    \end{align*}
    where $g_i(x) = W(x)|f_i(x) - F_i|$.
    The gradient of $\mathcal{L}_{IS}$ with respect to $\alpha$ can thus be obtained through the chain rule.

    \subsection{Alternative definitions of the objective function}
        \label{appendix:kl}
        \par As mentioned in the main text, one knows analytically that the distribution minimizing the variance of each gradient component is:
        \begin{equation*}
            q_{opt}^i(x) \propto |\psi(x)|^2(x)|f_i(x) - F_i|.
        \end{equation*}
        We will denote $q_{opt}^{SNIS}(x) = \frac{1}{N_p}\sum_{i=1}^{N_p} q_{opt}^i(x)$ the average of the above distributions, and $q_\alpha$ a variational family of distribution. One can choose a measure of divergence between probability distribution as the objective function, like the Kullback-Leibler divergence \cite{li_variance_2018}:
        \begin{equation*}
            \mathrm{D_{KL}}(q_\alpha |q_{opt}^{SNIS}) = \E_{q_\alpha}\left[\log\left(\frac{q_\alpha(x)}{q_{opt}^{SNIS}(x)}\right)\right].
        \end{equation*}
        One could also consider the following distribution, which is the average of the optimal distributions for standard importance sampling:
        \begin{equation*}
            q_{opt}^{IS}(x) \propto \sum_{i=1}^{N_p}|\psi(x)|^2(x)|f_i(x)|.
        \end{equation*}
        
\begin{figure}[ht]
            \center
            \includegraphics[width=\linewidth]{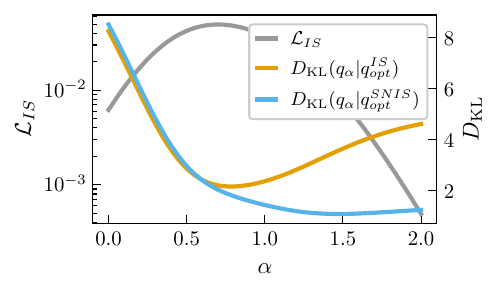}
            \caption{Comparison of several objective for the adaptive tuning. We compare Signal-to-noise ratio (grey) with Kullback Leibler divergence with respect to near optimal distributions for importance sampling (orange and blue). The comparison is carried on the state obtained at the end of the ground state search for the nitrogen molecule.}
            \label{fig:snr-kl}
        \end{figure}

        In \cref{fig:snr-kl}, we show that the minimums of KL divergences almost coincides with the maximum SNR for $q_{opt}^{IS}$ but not  $q_{opt}^{SNIS}$. Indeed, averaging all probability distribution may break optimality in a more dramatic way for $q_{opt}^{SNIS}$.

    \begin{figure}
            \center
            \includegraphics[width=\linewidth]{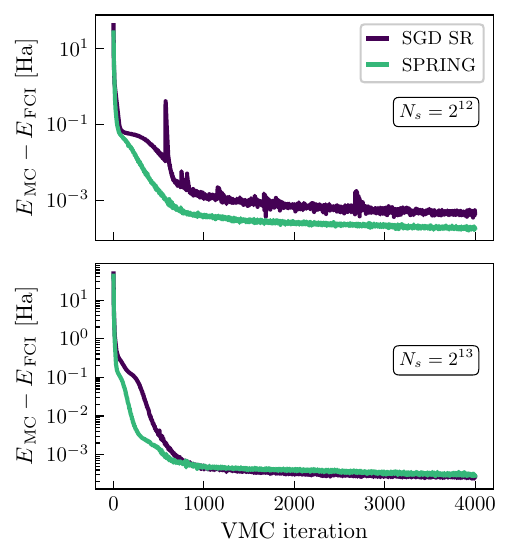}
            \caption{Comparison of SPRING with $\mu=0.9$ vs SGD SR using two different sample sizes. SPRING improves the optimization at low sample sizes.}
            \label{fig:springcomp}
    \end{figure}
    
        \section{Experimental details}
            \subsection{$J_1$-$J_2$ results}
                \subsubsection{ViT architecture}
                \label{appendix:vit}
                The Vision Transformer, as introduced in \cite{viteritti_transformer_2023}, is a deep neural network architecture originally applied to Computer Vision problems. First, one patches the input by grouping together nearby sites on the lattice. It is particularly interesting in the case of the square $J_1$-$J_2$ model as using patches of lateral size 2 corresponds to the unit cell of the model. Then, a linear embedding layer transforms these patches into $h$ vectors of dimension $d_{model}/h$, where $h$ denotes the number of heads and $d_{model}$ the total number of features. We will denote this matrix $X_i^\mu$, where the upper index runs on the number of heads. These vectors are then passed through a series of multi-head factored attention layers, whose central part is the following operation:
                \begin{equation*}
                    \alpha_i^\mu = \sum_{j=1}^{N} A_{i,j}^\mu X_j^\mu,
                    \label{facatt}
                \end{equation*}
                where $N$ denotes the number of patches, and $A_{i,j}^\mu$ is a tensor of variational parameters.

                Eq. \ref{facatt} essentially mixes the patches' embeddings, allowing the model to capture long-range correlations between them. The full ViT is a stack of these factored attentions combined with a patchwise Multi-Layer Perceptron to process the information aggregated by the factored attention in each patch vector. This architecture makes the ViT translational invariant on patch level. To impose full translational invariance, we just have to apply it at the level of the patch, by averaging the model over one-site translations. 
                
                This same architecture is also used for the SNR analysis on the 4x4 lattice, the only difference being that we employed a patch size of 1, which greatly improved the results for this lattice size.
                
                \subsubsection{Hyperparameters}
                \label{appendix:hparams}
                For training the ViT on the J1-J2 model, we used a fixed budget of 6000 iterations using Stochastic Reconfiguration, in which all simulations converged. To regularize the Quantum Geometric Tensor (QGT), we add a diagonal shift, initialized at $10^{-2}$ and decreased to $10^{-4}$ during the first 1000 steps using a cosine decay schedule. The learning rate is annealed in a similar manner, from $O(10^{-3})$ to $O(10^{-4})$. To remove bias, every variational energy is computed as an expectation values under $|\psi|^2$.

    \subsection{Ab-initio quantum chemistry}
        
        \subsubsection{Neural network backflow}
            We use an NQS with the same architecture as the one used in Ref.~\cite{luo_backflow_2019} and Ref.~\cite{liu_efficient_2025}. 
            For results on $\rm Li_2O$, we use 2 hidden layers, and 256 hidden units. 
            For $\rm N_2$, the number of hidden units is reduced to 16. 
            To improve the results, we also enforce spin flip symmetry.

            \subsubsection{Optimization}

            The Hamiltonians of $\rm N_2$ and $\rm Li_2O$ are expressed in the STO-3G basis and canonical HF molecular orbitals. 
            We observed that with MCMC, using natural orbitals from CCSD makes the optimization slightly more difficult.

            We employ SR or MinSR as an optimization technique depending on the number of variational parameters.
            We also found that SPRING with a momentum of 0.9 improves the training stability, as shown in figure \cref{fig:springcomp}.

            For lithium oxide, we train for 4000 iterations the learning is decayed linearly from $0.1$ to $0.001$ in the first 1000 iterations while the diagonal shift is also decreased linearly from $10^{-3}$ to $10^{-4}$ in 500 steps.
            These parameters are very conservative for many simulations, but could give the overall best results, especially for vanilla VMC using $|\psi(x)|^2$.

        \subsection{Infidelity minimization}
            We employ the same hyperparameters as appendix F of Ref.~\cite{gravina_neural_2024}. As an ansatz, we use a convolutional neural network with 4 layers, a fixed kernel size of 3, and 10 channels in each layer. The diagonal shift is set automatically with the PI controller developed in the same paper.

    \subsection{Equation summary (implementation aid)}
    \label{app:eq-summary-impl-aid}
    In this section we give all the necessary equations for the reader to implement the method presented in this paper. 
    For details about the notation and derivation, please refer to the main text and later appendices.

    \subsubsection{VMC}
    We aim at minimizing: 
    \begin{equation}
        \mathcal{L}(\theta) = \E_{p}[\ell(x)]
    \end{equation}
    for some local objective $\ell(x)$. To estimate the gradient $\bm F = \grad_{\theta} \mathcal{L}(\theta)$, we resort to:
    \begin{equation*}
         F_i = \E_{p}[\underbrace{2\,\Delta\partial_{\theta_i}\!\log \psi_\theta(x)\,\,\Delta \ell(x)}_{f_i(x)}],
    \end{equation*}
    where $\Delta A(x) \equiv A(x) - \E_{p}[A]$. 
    
    The self normalized importance sampling estimator of the gradient is:
    \begin{equation*}
        F_i = \E_{q}[\underbrace{W(x) f_i(x)}_{f_i^q(x)}],
    \end{equation*}
    where we define $W(x) = w(x)/\E_{q}[w(x)]$ and $w(x) = p(x)/q(x) = |\psi(x)|^2/q(x)$.
    In a similar manner, we estimate the Quantum Geometric Tensor (QGT) as:
    \begin{equation*}
         S_{ij} 
            = \Re{\E_{q}\!\left[W(x) \, \Delta\partial_{\theta_i}\!\log \psi_\theta(x)^*\,\,\Delta\partial_{\theta_j}\!\log \psi_\theta(x)\right]},
    \end{equation*}

    \subsubsection{Adaptive tuning}
    The variance of the gradient estimator can be estimated as:
    \begin{equation*}
        \V_q[f_i^q(x)] = \E_{q}\left[W^2(x)|f_i(x) - F_i|^2\right].
        \label{varis}
    \end{equation*}
    
    We aim at maximizing the following objective function:
    \begin{equation*}
        \mathcal{L}_{\rm IS}(q) = \frac{1}{N_p}\sum_{i=1}^{N_p} \sqrt{\frac{|\E_q[f_i^q(x)]|^2}{\V_q[f_i^q(x)]}} =  \frac{1}{N_p}\sum_{i=1}^{N_p} \mathrm{SNR}(f_i^q),
    \end{equation*}
    over the family of overdispersed sampling distributions:
    \begin{equation*}
            \qty{q_\alpha(x) = \frac{|\psi_\theta(x)|^\alpha}{Z_\alpha},\quad \alpha\geq 0}.
    \end{equation*}
    We use:
    \begin{equation*}
        \mathbf{\alpha}' = \mathbf{\alpha} + \eta \partial_\alpha\mathcal{L}_{\rm IS}(q_{\alpha}).
    \end{equation*}
    The gradient of the variance with respect to the variational parameters can be expressed as:
     \begin{multline*}
         \partial_{\alpha} \V_{q_\alpha}[f_i^{q_\alpha}(x)] = -\E_{q_\alpha}[\\\Delta_{q_\alpha}\partial_{\alpha_i}\log f_\alpha(x) W^2(x)|f_i(x) - F_i|^2],
    \end{multline*}
     where
     \begin{equation*}
         \Delta_{q_\alpha} \partial_\alpha \log f_\alpha(x) = \partial_\alpha \log f_\alpha(x) - \E_{q_\alpha}[\partial_\alpha \log f_\alpha(x)].
     \end{equation*}
     This leads to :
     \begin{equation*}
        \partial_{\alpha} \mathrm{SNR}(f_i^{q_\alpha}(x)) = \frac{1}{2} \partial_{\alpha} \V_{q_\alpha}[f_i^{q_\alpha}(x)] \frac{|\E_{q_\alpha}[f_i^{q_\alpha}(x)]|}{\V_{q_\alpha}[f_i^{q_\alpha}(x)]^{3/2}},
    \end{equation*}
    from which we build an estimator of $\partial_\alpha\mathcal{L}_{\rm IS}(q_{\alpha})$.

\end{document}